
%
%
\input phyzzx
\hoffset=0.2truein
\voffset=0.1truein
\hsize=6truein
\def\TITLEPAGE{\frontpagetrue}
\def\CALT#1{\hbox to \hsize{\tenpoint \baselineskip=12pt
        \hfil \vtop{
        \hbox{\strut CALT-68-#1}
        \hbox{\strut DOE RESEARCH AND}
        \hbox{\strut DEVELOPMENT REPORT}
	 \hbox{\strut hep-th/9308044}}}}
\def\CALTECH{
        \address{California Institute of Technology,
    Pasadena, CA 91125}}
\def\TITLE#1{\vskip.5in \centerline{\fourteenpoint#1}}
\def\AUTHOR#1{\vskip.2in \centerline{#1}}

\def\ABSTRACT#1{\vskip.2in \vfil \centerline
            {\twelvepoint \bf Abstract}
                     #1 \vfil}
\def\ENDTITLEPAGE{\vfil \eject \pageno=1}
\hfuzz=5pt
\tolerance=10000
\TITLEPAGE
\CALT{1880}
\TITLE{Complementarity in Wormhole Chromodynamics
\foot{This work supported in part
by the U.S. Department of Energy under Grant No.
DE-FG03-92-ER40701}}
\AUTHOR{Hoi-Kwong Lo\foot{\tt hkl@theory3.caltech.edu},
Kai-Ming Lee\foot{\tt kmlee@theory3.caltech.edu}, and
John Preskill\foot{\tt preskill@theory3.caltech.edu}}
\CALTECH
\ABSTRACT{The electric charge of a wormhole mouth and the magnetic flux
``linked'' by the wormhole are non-commuting observables, and so cannot be
simultaneously diagonalized.  We use this observation to resolve some puzzles
in wormhole electrodynamics and chromodynamics.  Specifically, we analyze the
color electric field
that results when a colored object traverses a wormhole, and we discuss the
measurement of the wormhole charge and flux using Aharonov-Bohm interference
effects.  We suggest that wormhole mouths may obey conventional quantum
statistics, contrary to a recent proposal by Strominger.}
\ENDTITLEPAGE
\eject

\noindent {\it Introduction}:
Many years ago, Wheeler\Ref\wheeler{J. A. Wheeler, Ann. Phys. (N. Y.) {\bf 2}
(1957) 604; J. A. Wheeler, {\it Geometrodynamics}, Academic Press, New York
(1962).} and Misner and Wheeler\Ref\misner{C. W. Misner and J. A.
Wheeler, Ann. Phys. (N. Y.) {\bf 2} (1957) 525.} proposed that electric field
lines trapped in the topology of a multiply-connected space might explain the
origin of electric charge.  Consider a three-dimensional space with a handle
(or ``wormhole'') attached to it, where the cross section of the wormhole is a
two-sphere.  On this space, the source-free Maxwell equations have a solution
with electric field lines caught inside the wormhole throat.  One mouth of the
wormhole, viewed in isolation by an observer who is unable to resolve the small
size of the mouth, cannot be distinguished from a pointlike electric charge.
Only when the observer inspects the electric field more closely, with higher
resolution, does she discover that the electric field is actually source free
everywhere.

It is also interesting to consider what happens when a charged particle
traverses a wormhole.\foot{Note that we are assuming that the wormhole is
traversable, in violation of the ``topological censorship''
theorem\Ref\friedman{J. Friedman, K. Schleich, and D. Witt, ``Topological
censorship,'' ITP preprint NSF-ITP-93-80 (1993).} that can be proved in
classical general relativity (with suitable assumptions about the positivity of
the energy-momentum tensor).  This traversability might be enforced by quantum
effects.  Alternatively, the reader might prefer to envision our space as a
thin two-dimensional film, containing objects with Aharonov-Bohm interactions.
Such wormholes might actually be fashioned in the laboratory!}
(Of course, this ``pointlike'' charge might actually be one mouth of a smaller
wormhole.)  Suppose that, initially, the mouths of the wormhole are uncharged
(no electric flux is trapped in the wormhole).  By following the electric field
lines, we see that after an object with electric charge $Q$ traverses the
wormhole, the mouth where it entered the wormhole carries charge $Q$, and the
mouth where it exited carries charge $-Q$.  Thus, an electric charge that
passes through a wormhole transfers charge to the wormhole mouths.

In this note, we will address two (closely related) puzzles associated with
this type of charge transfer process.  Our first puzzle concerns the quantum
mechanics of charged particles in the vicinity of a wormhole.  We can compute
the amplitude for the particle to propagate from an initial position to a final
position by performing a sum over histories.  Naively, one would expect this
sum to include histories that traverse the wormhole, and that the contribution
to the path integral due to these histories should be combined coherently with
the contribution due to histories that do not traverse the wormhole.  In fact,
the histories can be classified according to their ``winding number'' around
the wormhole, which can take any integer value, and one expects that all of the
winding sectors should be combined coherently.  Upon further reflection,
though, one sees that, for charged particles, this naive expectation must be
incorrect.  Long after the final position of the particle has been detected, an
observer can measure the charge of one of the wormhole mouths.  If the mouth
was uncharged initially, and carries charge $nQ$ finally, then the observer
concludes that the charged particle must have entered that mouth of the
wormhole $n$ times.  Because the winding sectors are perfectly correlated with
the charge transferred to the mouth, the amplitudes associated with different
numbers of windings cannot interfere with one another.  The puzzle in this case
is to understand more clearly the mechanism that destroys the coherence of the
different winding sectors.

Our second puzzle arises in a non-abelian gauge theory, such as quantum
chromodynamics.  Suppose that a wormhole initially carries no color charge, and
consider what happens when a ``red'' quark traverses the wormhole.  (We can
give a gauge-invariant meaning to the notion that the quark is red by
establishing a ``quark bureau of standards'' at some preferred location, and
carefully preserving a standard red ($R$) quark, blue ($B$) quark, and yellow
($Y$) quark there.  When we say that a quark at another location is red, we
mean that if it is parallel transported back to the bureau of standards, its
color matches that of the standard $R$ quark.  This notion is especially simple
if we assume that there are no color magnetic fields, so that parallel
transport is unaffected by smooth deformations of the path.)  An observer who
watches the red quark enter
one mouth of the wormhole concludes that the mouth becomes a red source of
color electric field.\foot{We are assuming that the wormhole is being examined
on a sufficiently short distance scale that the effects of color confinement
can
be neglected.} But the other mouth of the wormhole is initially in a
color-singlet state, and it cannot suddenly acquire a long-range color electric
field as the quark emerges from the mouth.  Thus, after the traversal, the
quark and mouth must be in the color-singlet state
$$
{1\over\sqrt{3}}\left(\ket{R}_{\rm quark}\otimes\ket{\bar R}_{\rm mouth}+
\ket{B}_{\rm quark}\otimes\ket{\bar B}_{\rm mouth}+ \ket{Y}_{\rm
quark}\otimes\ket{\bar Y}_{\rm mouth}\right)\, .
\eqn\singlet
$$
The puzzle in this case is to understand why the quark that emerges from the
wormhole is not simply in the color state $R$, and how the correlation between
the color of the quark and the color of the mouth is established.

The resolution of these puzzles involves some peculiar features of the
Aharonov-Bohm effect\Ref\AB{Y. Aharonov and D. Bohm, Phys. Rev. {\bf 119}
(1959)
485.} on non-simply connected manifolds.  The essential concept is the magnetic
flux ``linked'' by the wormhole.  If a particle with charge $Q$ is carried
around a closed path that traverses a wormhole (in a $U(1)$ gauge theory), it
in general acquires an
Aharonov-Bohm phase $e^{iQ\Phi}$, where $\Phi$ is the flux associated with the
path.  (This flux is defined modulo the flux quantum $\Phi_0=2\pi/e$, where $e$
is the charge quantum.)  If magnetic field strengths vanish everywhere, this
flux is a topological invariant, unchanged by smooth deformations of the path.
The crucial point is that the flux $\Phi$ and the charge of a wormhole mouth
are complementary observables---if the mouth has a definite charge (like zero),
then the flux does not take a definite value.  Summing over the different
possible values of the flux generates the decoherence of the winding sectors
described above, and also (in the non-abelian case) causes the red quark that
traverses the wormhole to
emerge in the state Eq.~\singlet.

In the concluding portion of the paper, we address a tangentially related
issue, the quantum statistics obeyed by identical wormhole mouths.

\medskip
\noindent{\it Wormhole complementarity}:
Let us now analyze these Aharonov-Bohm interactions in greater detail.  We will
use a notation that is appropriate when the gauge group $G$ is a finite group.
This will serve to remind the reader that our analysis applies to the case of a
{\it local discrete symmetry}.\REFS\krausswil{L. Krauss and F. Wilczek, Phys.
Rev. Lett. {\bf 62} (1989) 1221.}\REF\preskrauss{J. Preskill and L. Krauss,
Nucl. Phys. B {\bf 341} (1990) 50.}\REF\almrwil{M. Alford, J. March-Russell,
and F. Wilczek, Nucl. Phys. B {\bf 337} (1990) 695.}\refsend  For the case of a
continuous gauge group, one need only replace sums by integrals in some of the
expressions below.  When the gauge group is discrete (and also when it is
continuous), the electric charge of an object, including a wormhole mouth, can
be measured in principle by scattering a loop of {\it cosmic string} (or a
closed magnetic solenoid) off of the object.  For ease of visualization, we
will carry out our explicit analysis for the case of two spatial dimensions, so
that charges are measured by scattering magnetic {\it vortices}.  The analysis
in three spatial dimensions is essentially identical.

There are actually two types of topological magnetic flux associated with a
wormhole, for there are two topologically distinct paths for which
Aharonov-Bohm phases can be measured, as shown in Fig.~1.  The path $\alpha$
encloses one mouth of the wormhole, and we will denote the group element
associated with parallel transport around this path as $a\in G$.  The path
$\beta$ passes through both wormhole mouths, and we denote the associated group
element as $b\in G$.  We refer to these group elements as the $\alpha$-flux and
$\beta$-flux of the wormhole, and denote the corresponding quantum state of the
wormhole as $\ket{a,b}_{\rm wormhole}$.

Now, we can measure the {\it electric} charge of a wormhole mouth by winding a
vortex around the mouth, and observing the Aharonov-Bohm phase acquired by the
vortex.  However, winding the vortex around the mouth will also change the
state $\ket{a,b}$ of the wormhole.  For our purposes, it will be sufficient to
consider the special case in which $a=e$, the identity.  (A much more general
analysis of non-abelian Aharonov-Bohm interactions on topologically nontrivial
spaces can found in \REF\lee{K.-M.~Lee, Vortices on Higher Genus
Surfaces, Caltech preprint CALT-68-1873 (1993).} Ref.~\lee).  As shown in
Fig.~2, we may enclose the vortex with a closed path $\gamma$; we denote the
group element associated with transport around $\gamma$ as $h\in G$, and refer
to it as the flux of the vortex.  As the vortex winds counterclockwise around
the wormhole mouth, the path $\beta\gamma^{-1}$ is deformed to $\beta$.  (Here,
$\beta\gamma^{-1}$ denotes the path that is obtained by tracing $\gamma^{-1}$
{\it first}, followed by $\beta$.)  Thus,
when the vortex winds around the mouth, the flux associated with
$\beta\gamma^{-1}$ {\it before} the winding becomes the flux associated with
$\beta$ after the winding; we conclude that the state of wormhole and vortex is
modified according to
$$
\ket{e,b}_{\rm wormhole}\otimes\ket{h}_{\rm vortex}\to
\ket{e,bh^{-1}}_{\rm wormhole}\otimes\ket{h}_{\rm vortex}\, .
\eqn\windaround
$$
Eq.~\windaround\ is the centerpiece of our analysis.  It says that if the
wormhole is in the ``flux eigenstate'' $\ket{e,b}$, then any attempt to use
Aharonov-Bohm interference to measure the electric charge of one mouth is
doomed to failure.  If we scatter a vortex off of the mouth (with vortex flux
$h\ne
e$), whether the vortex passed to the left or the right of the mouth is
perfectly correlated with the state of the wormhole, and therefore no
interference is seen; the probability distribution of the scattered vortex is
the incoherent sum of the probability distributions for vortices that pass to
the left and pass to the right.

However, by superposing the wormhole states of definite $\beta$-flux, we can
construct
states with definite charge.  (We need only decompose the regular
representation of $G$ into irreducible representations.)  In particular, in the
state
$$
\ket{0}_{\rm wormhole}={1\over \sqrt{n_G}}\sum_{b\in G}\ket{e,b}_{\rm wormhole}
\eqn\zeromouth
$$
(where $n_G$ is the order of the group $G$), each mouth of the wormhole has
zero charge.  To see this, consider carrying the $h$-vortex around one mouth of
this wormhole.  It is easy to see that the state of the wormhole is unmodified,
so that the Aharonov-Bohm phase acquired by the vortex is trivial.  On the
other hand, suppose that we try to measure the $\beta$-flux of the wormhole by
carrying
a charged particle along the path $\beta$.  Let us denote the initial state of
the particle as $\ket{v}_{\rm particle}$, and let $(\nu)$ be the irreducible
representation of $G$ according to which the state transforms.  Then if we
carry this particle around the path $\beta$ where the wormhole is initially in
the state $\ket{0}_{\rm wormhole}$, the state of particle and wormhole is
modified according to
$$
\eqalign{&\ket{\rm initial}\equiv\ket{v}_{\rm particle}\otimes\ket{0}_{\rm
wormhole}\to\cr &\ket{\rm final}\equiv
{1\over \sqrt{n_G}}\sum_{b\in G}D^{(\nu)}(b)\ket{v}_{\rm particle}
\otimes\ket{e,b}_{\rm wormhole}\, ;\cr}
\eqn\betaaround
$$
thus the overlap of the final state with the initial state is
$$
\langle{\rm final}\ket{\rm initial}={1\over n_G}\sum_{b\in G}
\bra{v}D^{(\nu)}(b)\ket{v}
=\cases{1,&if $(\nu)={\rm trivial}\,$ ;\cr
0, &${\rm otherwise}\,$ .\cr}
\eqn\trivinterfer
$$
Unless $(\nu)$ is trivial, the state of the particle that has been carried
through the wormhole is orthogonal to the original state.
Hence we recover our earlier conclusion that, for  charged particles
propagating on the wormhole geometry, paths that traverse the wormhole add
incoherently with paths that do not.

We see that the wormhole cannot simultaneously have a definite $\beta$-flux and
a
definite charge.  We call this property ``wormhole complementarity.''  It is
intimately related to the complementary connection between magnetic and
electric flux that was first emphasized by 't Hooft,\Ref\hooft{G. 't Hooft,
Nucl. Phys. B {\bf 138} (1978) 1; Nucl. Phys. B {\bf 153} (1979) 141.} and was
generalized to the non-abelian case in \REF\ALMP{M. Alford, K.-M. Lee, J.
March-Russell, and J. Preskill, Nucl. Phys. B {\bf 384} (1992) 251; M. Bucher,
K.-M. Lee, and J. Preskill, Nucl. Phys. B {\bf 386} (1992) 27.}Ref.~\ALMP.

By decomposing the regular representation Eq.~\windaround\ into irreducible
representations, we obtain states in which the wormhole {\it mouth} has a
definite charge.  The charge of a mouth should not be confused with the
``Cheshire charge''\REF\coleman{M. G. Alford,
K. Benson, S. Coleman, J. March-Russell
and F. Wilczek, Phys. Rev. Lett. {\bf 64} (1990) 1632; Nucl. Phys. B
{\bf 349} (1991) 414.}\refmark{\preskrauss,\coleman} carried by the whole
wormhole.  To measure the charge of the whole wormhole, we would wind a vortex
around {\it both} mouths of the wormhole.  In this process, the state of vortex
and wormhole is modified according to\refmark{\lee}
$$\eqalign{
&\ket{a,b}_{\rm wormhole}\otimes\ket{h}_{\rm vortex}\to\cr
&\ket{hah^{-1}, hbh^{-1}}_{\rm
wormhole}\otimes\ket{h\left(aba^{-1}b^{-1}\right)h
\left(aba^{-1}b^{-1}\right)^{-1} h^{-1}}_{\rm vortex}\, .\cr}
\eqn\cheshire
$$
Note that $aba^{-1}b^{-1}$ is the ``total flux'' of the wormhole, the flux
associated with a path that encloses both mouths.  Charge measurement is
possible only if the initial and final vortex states are not orthogonal, so
that interference can occur.  Therefore, the flux $h$ of the vortex must
commute with the total flux of the wormhole---the charge that can be detected
is actually a representation of $N(aba^{-1}b^{-1})$, the centralizer of the
total flux.\REF\bala{A. P. Balachandran, F. Lizzi, and V. Rogers, Phys. Rev.
Lett. {\bf 52} (1984) 1818.}\refmark{\bala,\preskrauss,\coleman}  States of
definite Cheshire charge are obtained by decomposing the wormhole states
$\ket{a,b}$ into states that transform irreducibly under the action
Eq.~\cheshire, where $h\in N(aba^{-1}b^{-1})$.

Of course, to an observer with poor resolution, the wormhole mouths look like
pointlike particles, and the Cheshire charge of the wormhole coincides with the
Cheshire charge of vortex pairs that has been discussed
elsewhere.\REF\disentangling{M. Alford, S. Coleman, and J. March-Russell, Nucl.
Phys. B {\bf 351} (1991) 735.}\REF\lo{H.-K. Lo and J. Preskill, ``Non-abelian
vortices and non-abelian statistics,'' Caltech preprint CALT-68-1867
(1993).}\refmark{\ALMP,\disentangling-\lo}  For example, if $b=e$  then the
mouths appear to be a vortex with flux $a$ and an anti-vortex with flux
$a^{-1}$.  In the case $a=e$ that we have considered, neither wormhole mouth
carries any flux, and the states $\ket{e,b}_{\rm wormhole}$ are transformed as
$$
\ket{e,b}_{\rm wormhole}\otimes\ket{h}_{\rm vortex}\to
\ket{e,hbh^{-1}}_{\rm wormhole}\otimes\ket{h}_{\rm vortex}
\eqn\cheshnoflux
$$
when the vortex winds around the wormhole.  The states of definite Cheshire
charge are obtained by superposing the flux eigenstates $\ket{e,b}_{\rm
wormhole}$, with $b$ taking values in a particular conjugacy class of $G$.
Specifically the states
$$
\ket{0,[b]}_{\rm wormhole}={1\over \sqrt{n_{[b]}}}\sum_{b'\in
[b]}\ket{e,b'}_{\rm wormhole}
\eqn\nocheshire
$$
(where $[b]$ denotes the class containing $b$, and $n_{[b]}$ is the order of
that
class) have trivial total charge, although each wormhole mouth carries charge
in these states.

The peculiar behavior we found for Aharonov-Bohm scattering off of a wormhole
mouth, when the wormhole is in a flux eigenstate, can be given a more
conventional interpretation if we think of the wormhole as a pair of charged
particles in a particular (correlated) state.  For example, the flux eigenstate
$\ket{e,e}_{\rm wormhole}$ can be decomposed as
$$
\eqalign{
\ket{e,e}_{\rm wormhole}&\equiv\ket{0,[e]}_{\rm wormhole}\cr =&\sum_\nu
C_\nu\sum_i{1\over \sqrt{n_\nu}}
\ket{e_i,\nu}\otimes\ket{e_i^*,\nu}\, ,\quad
\sum_\nu|C_\nu|^2=1\, ,\cr}
\eqn\nocheshdecomp
$$
where the $\ket{e_i,\nu}$'s are a basis for the space on which the irreducible
representation $(\nu)$ acts, and $n_\nu$ is the dimension of this
representation.
This is a superposition of states in which the two particles (the
mouths) have nontrivial charges, and are in a combined state of trivial charge.
Experiments involving one of the mouths are described by a mixed
density matrix of the form
$$
\rho=\sum_\nu|C_\nu|^2{1\over n_\nu} {\bf 1}_\nu\, ,
\eqn\density
$$
and Aharonov-Bohm scattering of the $h$-vortex off the mouth enables us to
measure
$$
{\rm tr}~D(h)\rho=\sum_\nu|C_\nu|^2{1\over n_\nu}\chi^{(\nu)}(h)
=\cases{1, &$h=e\,$ ;\cr 0, &${\rm otherwise}\,$ ,\cr}
\eqn\ABfluxeigen
$$
where $\chi^{(\nu)}$ denotes the character of the representation.  (The second
equality in Eq.~\ABfluxeigen\ follows from the property Eq.~\windaround.)  From
the group orthogonality relations, we see that $|C_\nu|^2=n_\nu^2/n_G$.  Thus
Aharonov-Bohm scattering enables us to determine the probability that the
wormhole mouth carries charge $(\nu)$, but does not determine the relative
phases of the $C_\nu$'s.\refmark{\disentangling}  When we think of it as a
point particle, the unusual thing about a wormhole mouth is that it is natural
to consider a state such that the mouth is in a
superposition of particle states with different gauge charges.

\medskip
\noindent{\it Charge transfer}:
Now let us suppose that, after the wormhole in the initial state $\ket{0}_{\rm
wormhole}$ is traversed by the charged particle in the initial state
$\ket{v}_{\rm particle}$, we attempt again to measure the charges of the two
mouths.  If an $h$-vortex is carried around the mouth that the charged particle
{\it entered}, then the state of wormhole, particle, and vortex is modified
according to
$$
\eqalign{&{1\over \sqrt{n_G}}\sum_{b\in G}D^{(\nu)}(b)\ket{v}_{\rm
particle}\otimes \ket{e,b}_{\rm wormhole}\otimes \ket{h}_{\rm vortex}\to\cr
&{1\over \sqrt{n_G}}\sum_{b\in G}D^{(\nu)}(b)\ket{v}_{\rm particle}\otimes
\ket{e,bh^{-1}}_{\rm wormhole}\otimes \ket{h}_{\rm vortex}\, ,}
\eqn\mouthagain
$$
so that the overlap of the initial state with the final state is
$$
{\rm overlap}={1\over n_G}\sum_{b,b'\in
G}\bra{v}D^{(\nu)}(b')^{-1}D^{(\nu)}(b)\ket{v} \cdot
\bra{e,b'}e,bh^{-1}\rangle=\bra{v}D^{(\nu)}(h)\ket{v}\, .
\eqn\samAB
$$
This is exactly the same as the overlap we would have obtained if the vortex
had been carried around the initial charged particle.  Thus, as we anticipated,
the charge of the particle has been transferred to the mouth of the wormhole.

But if we measure instead the charge of the {\it other} mouth, we obtain a
rather different result.  It is actually most instructive to consider carrying
the $h$-vortex around {\it both} the charged particle and the other wormhole
mouth.  A variant of the argument given earlier shows that carrying the vortex
counterclockwise around this mouth changes the wormhole state $\ket{e,b}$ to
$\ket{e, hb}$.  We thus find that the state of wormhole, particle, and vortex
is modified according to
$$
\eqalign{&{1\over \sqrt{n_G}}\sum_{b\in G}D^{(\nu)}(b)\ket{v}_{\rm
particle}\otimes \ket{e,b}_{\rm wormhole}\otimes \ket{h}_{\rm vortex}\to\cr
&{1\over \sqrt{n_G}}\sum_{b\in G}D^{(\nu)}(h)D^{(\nu)}(b)\ket{v}_{\rm
particle}\otimes \ket{e,hb}_{\rm wormhole}\otimes \ket{h}_{\rm vortex}\, ,}
\eqn\othermouth
$$
and that the overlap of the initial state with the final state is
$$
{\rm overlap}={1\over n_G}\sum_{b,b'\in
G}\bra{v}D^{(\nu)}(b')^{-1}D^{(\nu)}(hb)\ket{v} \cdot \bra{e,b'}e,hb\rangle=1\,
{}.
\eqn\samAB
$$
Thus the Aharonov-Bohm phase is trivial, and we conclude that the charged
particle and
mouth are combined together into a singlet state, again as anticipated.

Eq.~\singlet\ is a special case of this result.  We now understand that if the
wormhole mouth initially carries no color charge, that means that the color
holonomy associated with traversing the wormhole does not take a definite
value.  Thus the red quark emerges from the wormhole mouth carrying indefinite
color, but with its color perfectly anti-correlated with the color of the
mouth.  Furthermore, after the (initially) red quark passes through the
wormhole, the wormhole state is a superposition of a color octet and color
singlet, so that Cheshire charge has been transferred to the
wormhole.\foot{$SU(3)_{\rm color}$ Cheshire charge has also been discussed by
Bucher and Goldhaber.\Ref\buchgold{M. Bucher and A. S. Goldhaber, ``$SO(10)$
cosmic strings and $SU(3)_{\rm color}$ Cheshire charge,'' Inst. Adv. Study
preprint (1993).}}

\medskip
\noindent{\it Quantum statistics}:
The quantum-mechanical scattering of wormhole mouths was recently discussed by
Strominger.\Ref\strominger{A. Strominger, ``Black hole statistics,'' ITP
preprint NSF-ITP-93-57 (1993).}  He argued that ``identical'' wormhole mouths
behave differently than ordinary indistinguishable particles---they are neither
bosons nor fermions.  The idea is that an exchange of two mouths, with the
corresponding ``anti-mouths'' fixed, does not leave the geometry unchanged;
hence there is no need for the wave function to have any particular symmetry
properties under the interchange.  On the other hand, the mouths are not
exactly distinguishable particles either, because an exchange of two wormholes
(a simultaneous exchange of two mouths and the corresponding anti-mouths) {\it
does} preserve the geometry.

Strominger's wormhole mouths that obey neither Bose or Fermi statistics are
somewhat reminiscent of the wormhole mouths, discussed here, that have an
indefinite gauge charge.  Since it is possible to construct charge-eigenstate
mouths by superposing the wormhole flux eigenstates, this analogy suggests an
alternative approach to the quantum statistics of wormhole mouths.  For
example, in the case of two wormholes, we can consider superposing the two
possible ways of connecting mouths and anti-mouths.  These states have definite
symmetry properties under mouth
exchange---the mouths behave like bosons for the superposition with a plus
sign, and like fermions for the superposition with the minus sign.
Furthermore, we might wish to assume that all local observables are indifferent
to the way mouths and anti-mouths are
connected.  This assumption is very natural in the case considered by
Strominger, where the wormhole mouths have event horizons (they are extremal
black holes), and the wormholes are not traversable.  Then we may argue that
the two possible ways of quantizing the mouths---as bosons or
fermions---correspond to two distinct superselection sectors of the theory.
Note also that any instantons that allow the wormholes to ``intercommute'' will
preserve these sectors.

Strominger's picture is motivated by considering mouth--anti-mouth pair
production, for it seems natural that when a new pair is produced, the new
mouth
and antimouth are joined by a wormhole.  In contrast, the alternative view that
mouths are like ordinary bosons or fermions requires for consistency that when
a pair is produced, the new mouth is connected in all possible ways to the
existing anti-mouths. (Any one of the existing wormholes can ``break'' into two
wormholes when the new pair nucleates.)  From this perspective, the pair
production process
appears to be highly non-local.  However (as in related discussions of
wormholes in Euclidean spacetime\Ref\coleman{S. Coleman, Nucl. Phys. B {\bf
307} (1988) 867; {\bf 310} (1988) 643; S. Giddings and A. Strominger, Nucl.
Phys. B {\bf 307} (1988) 854.}), this non-locality may be {\it so} extreme that
the physics admits a local description.  Indeed, bosons and fermions can be
described by local quantum field theory, while no such
description is known for the wormhole mouths envisioned by Strominger.

\bigskip
We thank Patrick McGraw, Sandip Trivedi and Piljin Yi for helpful
conversations.  We are also very grateful to Andy Strominger for a careful
reading of the manuscript.
\refout

\FIG\figA{Two non-contractible paths $\alpha$ and $\beta$, beginning and ending
at an arbitrarily chosen basepoint $x_0$, on the wormhole geometry.  The group
elements associated with parallel transport around these paths are the
$\alpha$-flux and $\beta$-flux of the wormhole.}

\FIG\figB{A vortex winds around one mouth of the wormhole, as shown in (a).  If
the path $\beta\gamma^{-1}$ shown in (b) is deformed during the winding of the
vortex, so that the vortex never crosses the path, $\beta\gamma^{-1}$ evolves
to the path $\beta$.}

\figout
%
\bye